\documentclass[a4paper,11pt]{article}
\usepackage{pos}
\usepackage{tikz}
\usetikzlibrary{patterns}

\usepackage{graphicx}
\usepackage{caption}
\usepackage{subcaption}

\title{Symmetry Breaking in an Extended-O(2) Model}

\author*[a,b]{Leon Hostetler}
\author[c]{Ryo Sakai}
\author[d]{Jin Zhang}
\author[e]{Judah Unmuth-Yockey}
\author[b,a]{Alexei Bazavov}
\author[d]{Yannick Meurice}

\affiliation[a]{Department of Physics and Astronomy, Michigan State University, East Lansing, Michigan 48824, USA}
\affiliation[b]{Department of Computational Mathematics, Science and Engineering, Michigan State University, East Lansing, Michigan 48824, USA}
\affiliation[c]{Department of Physics, Syracuse University, Syracuse, NY 13210, USA}
\affiliation[d]{Department of Physics and Astronomy, The University of Iowa, Iowa City, Iowa 52242, USA}
\affiliation[e]{Fermilab, Batavia, Illinois 60510, USA}

\emailAdd{hostet22@msu.edu}

\abstract{Motivated by attempts to quantum simulate lattice models with continuous Abelian symmetries using discrete approximations, we consider an extended-O(2) model that differs from the ordinary O(2) model by an explicit symmetry breaking term. Its coupling allows to smoothly interpolate between the O(2) model (zero coupling) and a $q$-state clock model (infinite coupling). In the latter case, a $q$-state clock model can also be defined for non-integer values of $q$. Thus, such a limit can also be considered as an analytic continuation of an ordinary $q$-state clock model to non-integer $q$. The phase diagram of the extended-O(2) model in the infinite coupling limit was established in our previous work, where it was shown that for non-integer $q$, there is a second-order phase transition at low temperature and a crossover at high temperature. In this work, we investigate the model at finite values of the coupling using Monte Carlo and tensor methods. The results may be relevant for configurable Rydberg-atom arrays.}

\FullConference{%
The 39th International Symposium on Lattice Field Theory,\\
8th-13th August, 2022,\\
Rheinische Friedrich-Wilhelms-Universität Bonn, Bonn, Germany
}

\begin{document}
\maketitle
	
\section{Introduction}

For the quantum simulation of quantum field theories, one needs to discretize space and truncate the fields. For example, one could make a $\mathbb{Z}_q$ approximation of a continuous $U(1)$ symmetry. Given the limited qubits available, it is important to optimize the discretization procedure, and it may be be useful to have a family of models that allows one to continously interpolate among various possible discretizations.

We consider a class of extended-O(2) models by adding a term $\gamma \cos (q\varphi)$ to the classical $O(2)$ model. The continuously tunable parameters $q$ and $\gamma$ allows us to interpolate among various models. When $\gamma=0$, the symmetry-breaking term is turned off, and this is just the $O(2)$ model. When $\gamma=\infty$, the $O(2)$ symmetry is completely broken, and the spins take only discrete angles $\varphi=2\pi k/q$ for integer $k$. For finite values of $\gamma$, we can study the effect of the symmetry-breaking. The parameter $q$ allows us to tune from the Ising model ($q=2$), through the $q$-state clock models, and to the $XY$ model ($q\rightarrow \infty$). We allow also noninteger $q$, which allows us to interpolate among all of these models. 

Previously, we studied the $\gamma\rightarrow\infty$ limit of this model \cite{Hostetler:2021uml,Hostetler:2021pos}, and we found that for noninteger $q$, there is a crossover at small $\beta$ and an Ising phase transition at large $\beta$. We present here some preliminary results for finite $\gamma$ by looking at the specific heat and entanglement entropy. These early results suggest that the $\gamma=\infty$ phase diagram for noninteger $q$ seems to persist to all $\gamma>0$. However, a rigorous determination of the phase diagram awaits the completion of our ongoing finite size scaling study of this model.

\section{The Extended-O(2) Model}

\begin{figure}
\centering
\begin{subfigure}{.24\textwidth}
  \centering
  \includegraphics[width=1\linewidth]{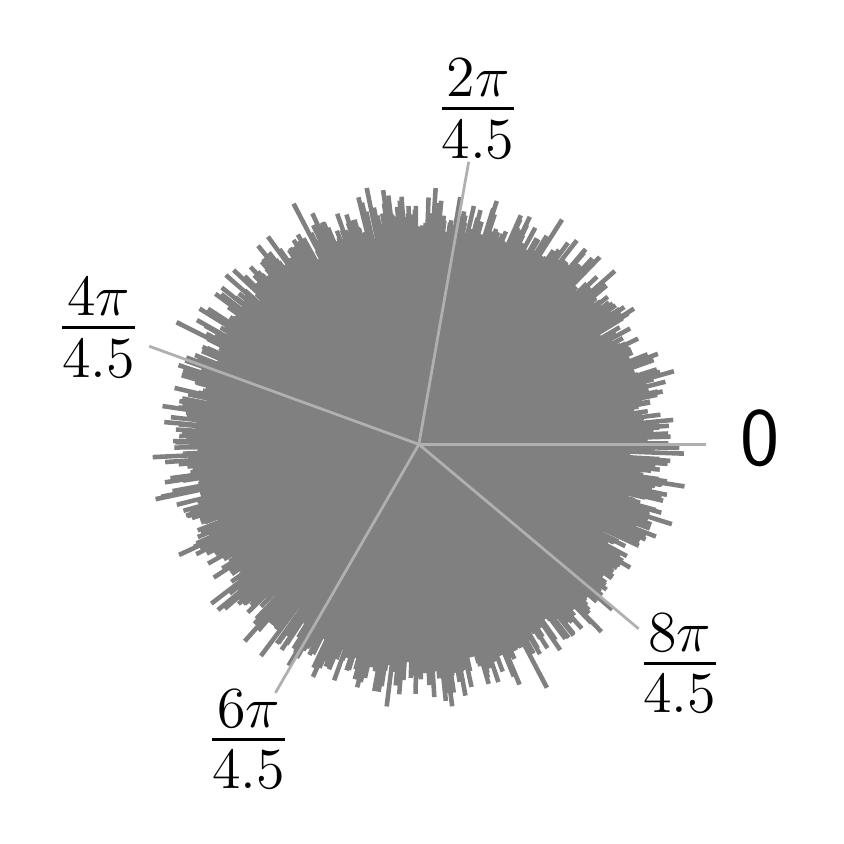}
  \label{fig:sub1}
\end{subfigure}%
\begin{subfigure}{.24\textwidth}
  \centering
  \includegraphics[width=1\linewidth]{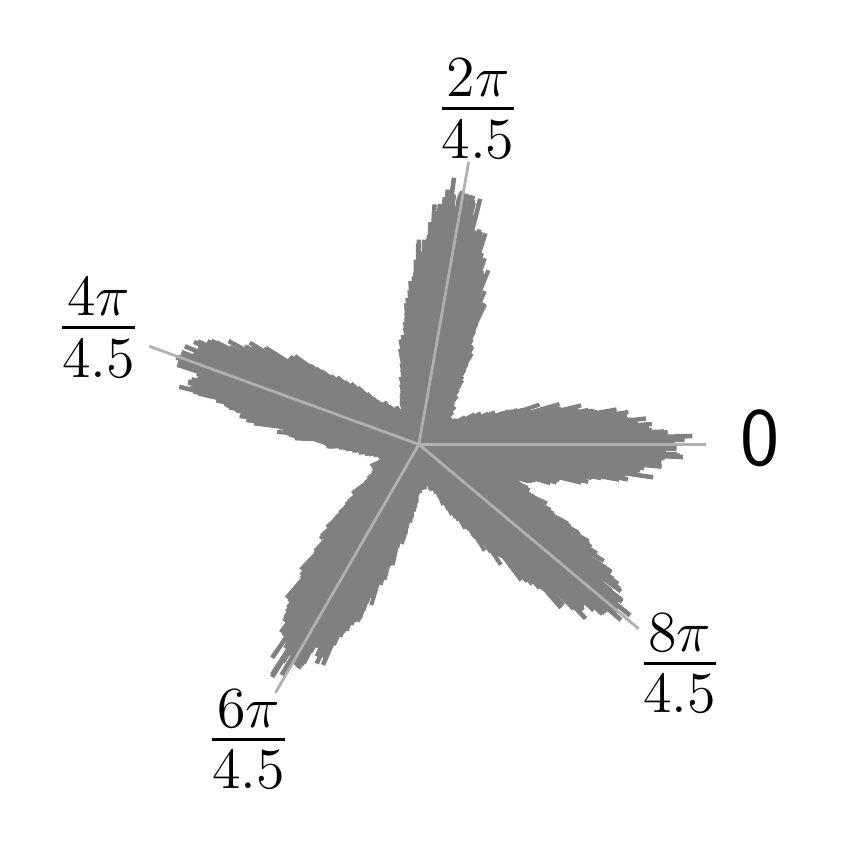}
  \label{fig:sub2}
\end{subfigure}
\label{fig:test}
\begin{subfigure}{.24\textwidth}
  \centering
  \includegraphics[width=1\linewidth]{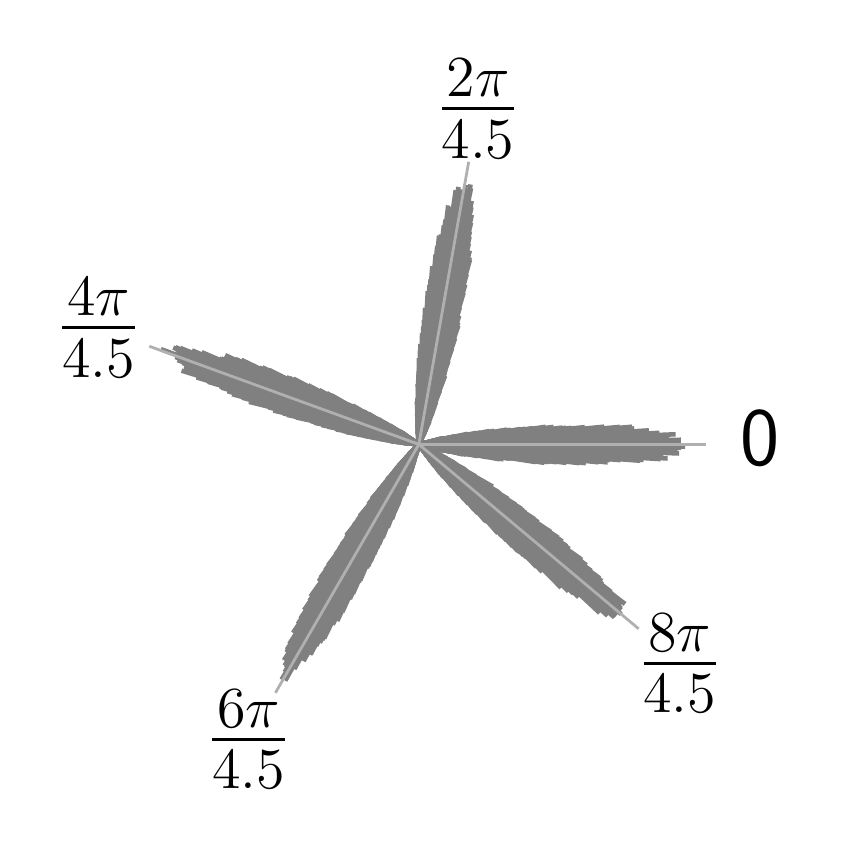}
  \label{fig:sub2}
\end{subfigure}
\label{fig:test}
\begin{subfigure}{.24\textwidth}
  \centering
  \includegraphics[width=1\linewidth]{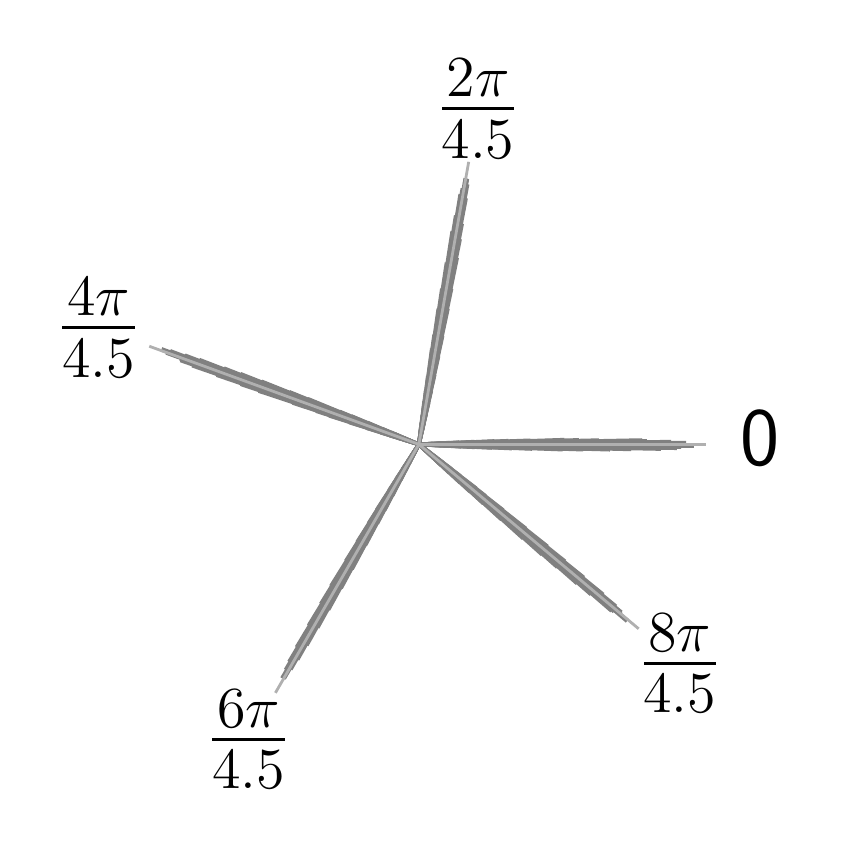}
  \label{fig:sub2}
\end{subfigure}
\caption{Here we illustrate how the spin angle distribution changes with $\gamma$ for the example $q=4.5$ and $\beta=0$. From left to right, we have $\gamma=0,1,4,64$. For large values of $\gamma$, the spins strongly prefer the ``clock'' angles defined by $\varphi = 2 \pi k/q$ for $k = 0,1,\ldots, \lfloor q\rfloor$. Note the ``leftover'' angle, which in this case is $\tilde{\phi} = \pi/4.5$.}
\label{fig:clock_angles}
\end{figure}

We study a two-dimensional classical spin model obtained by adding a symmetry-breaking term to the action
of the classical O(2) model
\begin{equation}
S_{\mbox{\scriptsize ext-}O(2)} = -\beta \sum_{x,\mu}\cos(\varphi_{x+\hat\mu}-\varphi_{x}) - \gamma\sum_x\cos(q\varphi_x).
\end{equation}
When $\gamma = 0$, this is the classic XY model, which is known to have a BKT transition. When $\gamma > 0$, the second term breaks periodicity and we must choose $\varphi \in[\varphi_0,\varphi_0 + 2\pi)$ for some choice $\varphi_0$. In this work, we use\footnote{The choice $\varphi \in [0,2\pi)$ results in a hard cutoff at $\varphi=0$. In practice, for noninteger $q$, we found that this cutoff skews the distribution of angles severely enough that the model at finite $\gamma$ does not smoothly connect to the model at infinite $\gamma$ when $\gamma\rightarrow\infty$. To fix this, we shifted the angle domain such that $\varphi \in [-\varepsilon, 2\pi-\varepsilon)$ to move the hard cutoff away from the clock angle at $\varphi=0$. We chose $\varepsilon = \pi(1-\lfloor q\rfloor/q)$.} $\varphi_0 = 0$. When $\gamma \rightarrow \infty$, the continuous angle $\varphi$ is forced into the discrete ``clock'' angles
\begin{equation}
\label{eq:clock_angles}
\varphi_0 \leq \varphi_{x,k}= \frac{2\pi k}{q} < \varphi_0 + 2\pi,
\end{equation}
as illustrated in Figure~\ref{fig:clock_angles}. For integer $q$, this is the ordinary $q$-state clock model with $\mathbb{Z}_q$ symmetry. For noninteger $q$, this defines an interpolation of the clock model to noninteger values of $q$. For noninteger $q$, divergence from ordinary clock model behavior is driven by the ``leftover'' angle
\begin{equation}
\label{eq:small_angle}
\tilde{\phi} = 2\pi\left(1-\frac{\lfloor q \rfloor }{q}\right).
\end{equation}

\begin{figure}
\centering
\includegraphics[scale=.7]{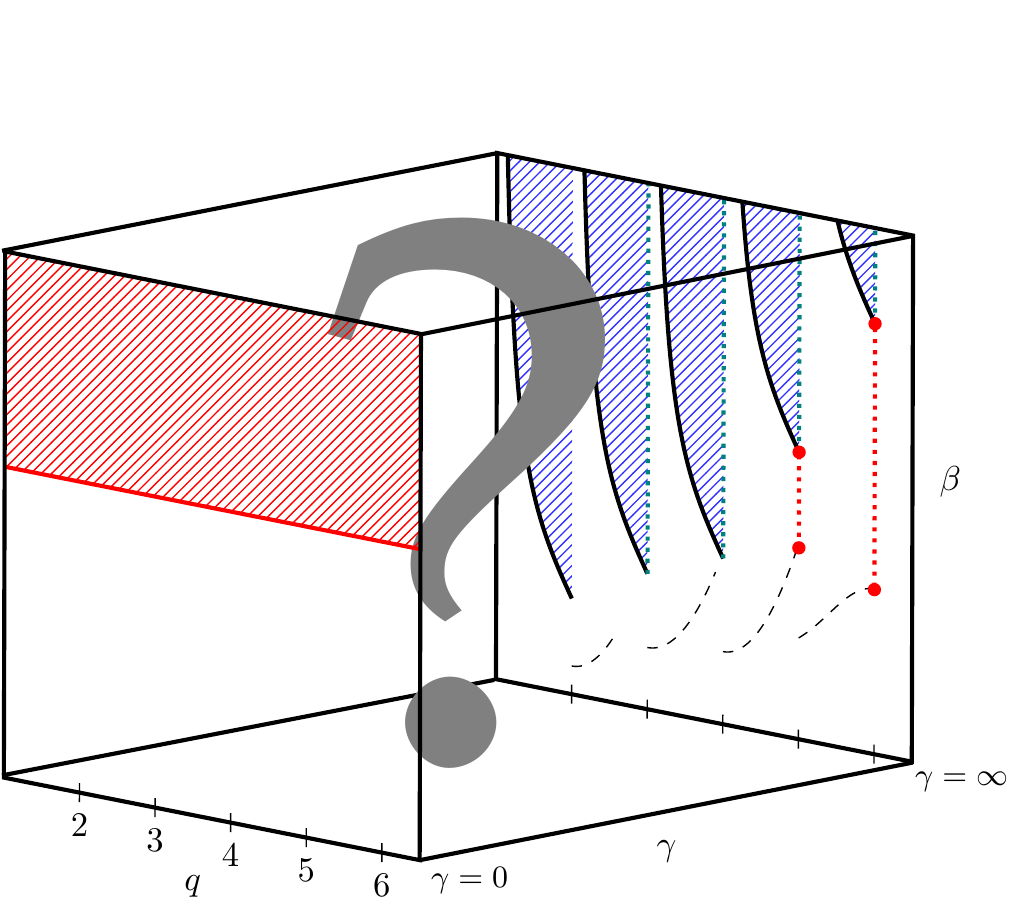}
\caption{In this Extended-O(2) model, the phase diagram is three-dimensional. In the $\gamma=0$ plane, it is the $XY$ model for all values of $q$. The $XY$ model has a disordered phase at small $\beta$, a BKT transition near $\beta_c = 1.12$, and a critical phase at large $\beta$. We previously established the phase diagram of the $\gamma=\infty$ plane \cite{Hostetler:2021uml}. For integer $q$, it is the well-studied clock model. For noninteger $q$, there is a crossover at small $\beta$ and a second-order phase transition of the Ising universality class at large $\beta$. Establishing the phase diagram at finite-$\gamma$ is the goal of the current and ongoing work.\label{fig:3dpd}}\end{figure}

Previously, we studied the $\gamma\rightarrow\infty$ limit of this model \cite{Hostetler:2021uml,Hostetler:2021pos}. In this limit, we were able to replace the action with
\begin{equation}
S_{\mbox{\scriptsize ext-}q} = -\beta\sum_{x,\mu}\cos(\varphi_{x+\hat\mu}-\varphi_{x}),
\end{equation}
but then directly restrict the previously continuous angles to the discrete values given by Eq.~(\ref{eq:clock_angles}). Using Markov Chain Monte Carlo (MCMC) and tensor renormalization group (TRG) methods, we were able to map out the $\gamma=\infty$ phase diagram of the model. For noninteger $q$, we found a crossover at small $\beta$ and a second-order phase transition of the Ising universality class at large $\beta$. We are currently working to understand the phase diagram at finite-$\gamma$. See Figure~\ref{fig:3dpd}.

We define the internal energy as
\begin{equation}
\label{eq_internalenergy}
\langle E\rangle= \left\langle -\sum_{x,\mu} \cos(\varphi_{x+\hat{\mu}} - \varphi_{x}) \right\rangle,
\end{equation}
and the specific heat as
\begin{equation}
\label{eq_specificheat}
C= \frac{-\beta^{2}}{V} 
\frac{\partial \langle E \rangle}{\partial \beta} = \frac{\beta^{2}}{V} (\langle E^{2} \rangle - \langle E \rangle ^{2}),
\end{equation}
where $\langle\dots\rangle$ denotes the ensemble average. 

The entanglement entropy is a quantum quantity, however, one can introduce an analog for classical lattice systems via a reduced density matrix where the degrees of freedom of the system are partially integrated out. On a tensor network this can be realized as a partial trace of indices. We define the entanglement entropy as
\begin{equation}
\label{eq_ee}
S_E = - \sum_i \rho_{A_i} \ln \left( \rho_{A_i} \right),
\end{equation}
where $\rho_{A_i}$ are the eigenvalues of a reduced density matrix $\hat{\rho}_A$. A detailed discussion can be found in \cite{yangetal}.

\section{Preliminary Results}

When we studied the model in the $\gamma\rightarrow \infty$ limit, we were able to treat the spin degrees of freedom as discrete. This allowed us to use an MCMC heatbath algorithm to explore the model at small volumes and a TRG method to study it at large volumes. The model is more difficult to study when $\gamma$ is finite. The spin degrees of freedom are now continuous. MCMC heatbath is no longer an option, so we're left with the Metropolis algorithm, which suffers from low acceptance rates and leads to large autocorrelations for noninteger $q$. Furthermore, our TRG method was designed for the $\gamma\rightarrow\infty$ limit, and extending it to finite-$\gamma$ proved difficult. We needed to make some algorithmic improvements. On the Monte Carlo side, we implemented a biased Metropolis heatbath algorithm \cite{bergbazavov:2005}, which is designed to approach heatbath acceptance rates. To explore large volumes, we implemented a tensorial approach that used Gaussian quadrature.

Recent previous work on a very similar\footnote{In our model, $\beta$ is treated as a coupling attached to the interaction term of the energy function, and then the Boltzmann factor is $e^{-S}$. In their model, $\beta$ is factored out, such that the Boltzmann factor is $e^{-\beta S}$, and $\beta$ therefore functions as a true thermodynamic inverse temperature. We believe the phase diagrams of these two models are qualitatively the same since the two models differ only by a rescaled $\gamma$.} model includes \cite{NoumanButt}. There, the authors studied integer $q \geq 5$ in the $\gamma\rightarrow 0$ limit as a symmetry-breaking perturbation to the $O(2)$ model. They found that a second transition (in addition to the BKT transition that occurs at $\gamma=0$) occurs for any finite $\gamma>0$. In the present work, we consider also intermediate and large values of $\gamma$ as well as noninteger values of $q$. Our initial exploration of the model at finite-$\gamma$ was performed using MCMC on small lattices, but then we used tensor methods to perform a large volume survey of the parameter space.

\begin{figure}
\centering
\includegraphics[scale=.6]{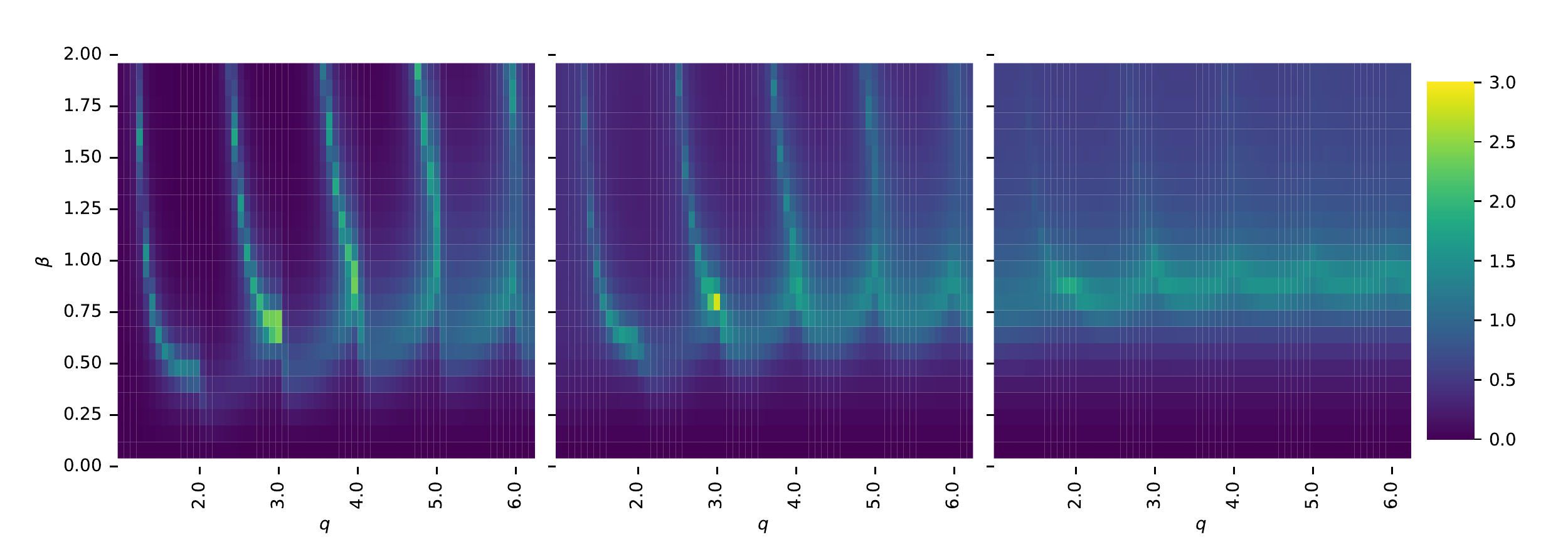}
\caption{Heatmaps of the specific heat from TRG with $L=1024$. \textbf{(Left panel)} Here $\gamma=16$. This already looks like the $\gamma=\infty$ phase diagram, where for noninteger $q$ there is a crossover at small $\beta$, and a second-order phase transition of the Ising universality class at large $\beta$. We see that there are discontinuities as $q$ crosses integer values. At large values of $\gamma$, the spin angles strongly prefer the ``clock angles'' $q=2\pi k/q$ as illustrated in Figure~\ref{fig:clock_angles}. As $q$ crosses an integer, an additional clock angle is introduced into the model, hence the discontinuites at integer $q$ when $\gamma$ is large. \textbf{(Middle panel)} Here $\gamma=1$. At this intermediate value of $\gamma$, the discontinuities have disappeared, and what remains are smooth curves. \textbf{(Right panel)} Here $\gamma=0.1$. At this small value of $\gamma$, the $O(2)$ symmetry is barely broken by the addition of the $\gamma \cos (q\varphi)$ term, and the specific heat ``phase diagram'' seems to be smoothly connecting to the $XY$ phase diagram at $\gamma=0$. \label{fig:cv_heatmaps}}
\end{figure}

In general, a plot of the specific heat versus temperature or $\beta$ shows a peak whenever there is phase transition or a crossover. Thus, a ``heatmap'' of the specific heat can serve as a proxy for the phase diagram. In Figure~\ref{fig:cv_heatmaps}, we show three such heatmaps of the specific heat in the $\beta$-$q$ plane for three different values of $\gamma$. These heatmaps were computed by TRG with $L=1024$ and $\Delta \beta = \Delta q = 0.08$. In the left panel, $\gamma=16$. This already looks like the $\gamma=\infty$ phase diagram, where for noninteger $q$ there is a crossover at small $\beta$, and a second-order phase transition of the Ising universality class at large $\beta$. We see that there are discontinuities as $q$ crosses integer values. At large values of $\gamma$, the spin angles strongly prefer the ``clock angles'' $q=2\pi k/q$ as illustrated in Figure~\ref{fig:clock_angles}. As $q$ crosses an integer, an additional clock angle is introduced into the model, hence the discontinuites at integer $q$ when $\gamma$ is large. In the middle panel, $\gamma=1$. At this intermediate value of $\gamma$, the discontinuities have disappeared, and what remains are smooth curves. In the right panel, $\gamma=0.1$. At this small value of $\gamma$, the $O(2)$ symmetry is barely broken by the addition of the $\gamma \cos (q\varphi)$ term, and the specific heat ``phase diagram'' seems to be smoothly connecting to the $XY$ phase diagram at $\gamma=0$.

In Figure~\ref{fig:cv_heatmaps}, we see that at intermediate values of $\gamma$, a heatmap of the specific heat develops smooth curves. This suggests the possibility of a continuous line of phase transitions even across integer values of $q$, or perhaps even of more exotic phases. For example, the heatmap of the specific heat at $\gamma=1$ shows similarities to the phase diagram of certain Rydberg atom systems \cite{Keesling:2019}, which also involve energy functions with continuously tunable parameters.

\begin{figure}
\centering
\begin{subfigure}{.5\textwidth}
  \centering
  \includegraphics[width=1\linewidth]{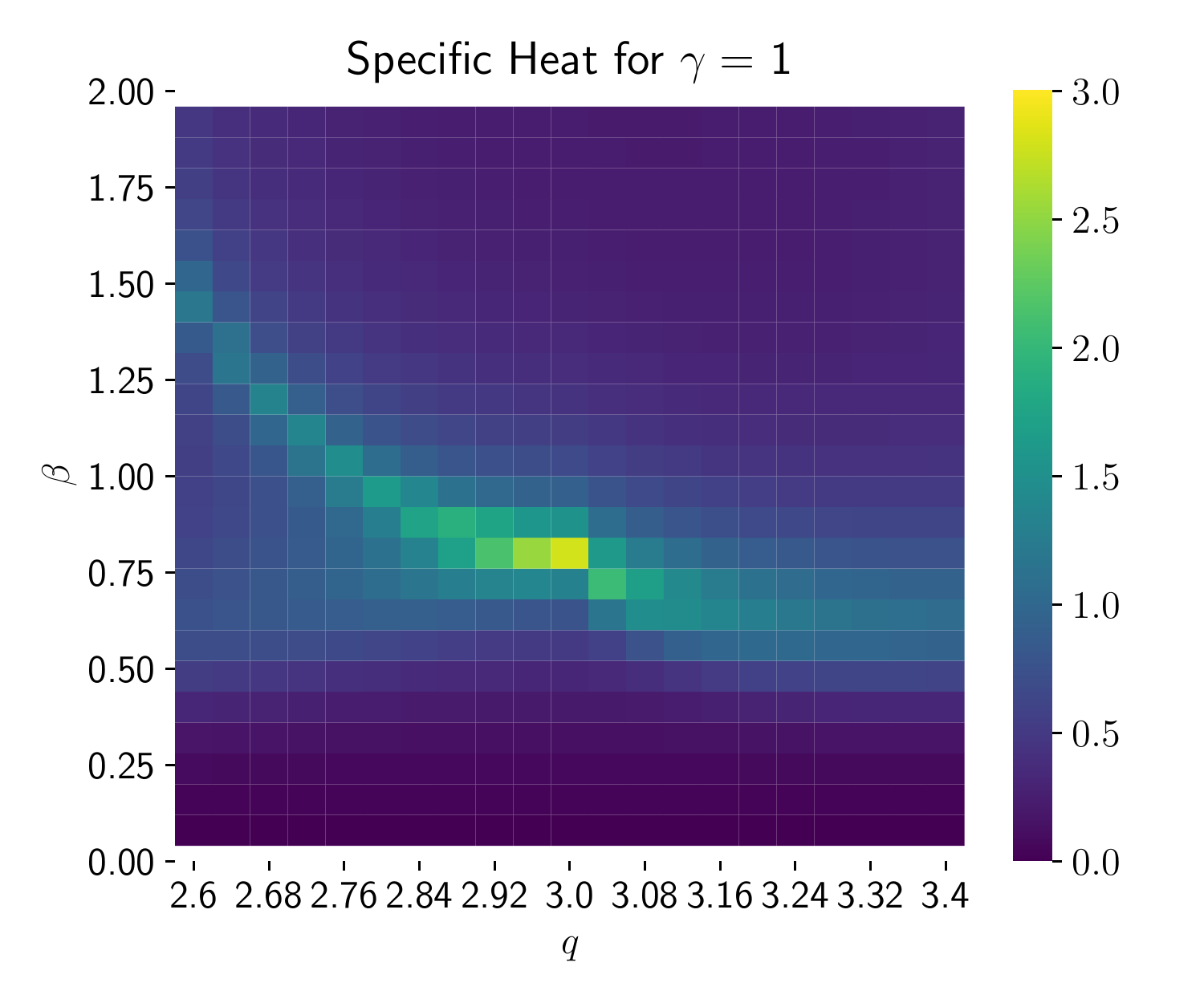}
\end{subfigure}%
\begin{subfigure}{.5\textwidth}
  \centering
  \includegraphics[width=1\linewidth]{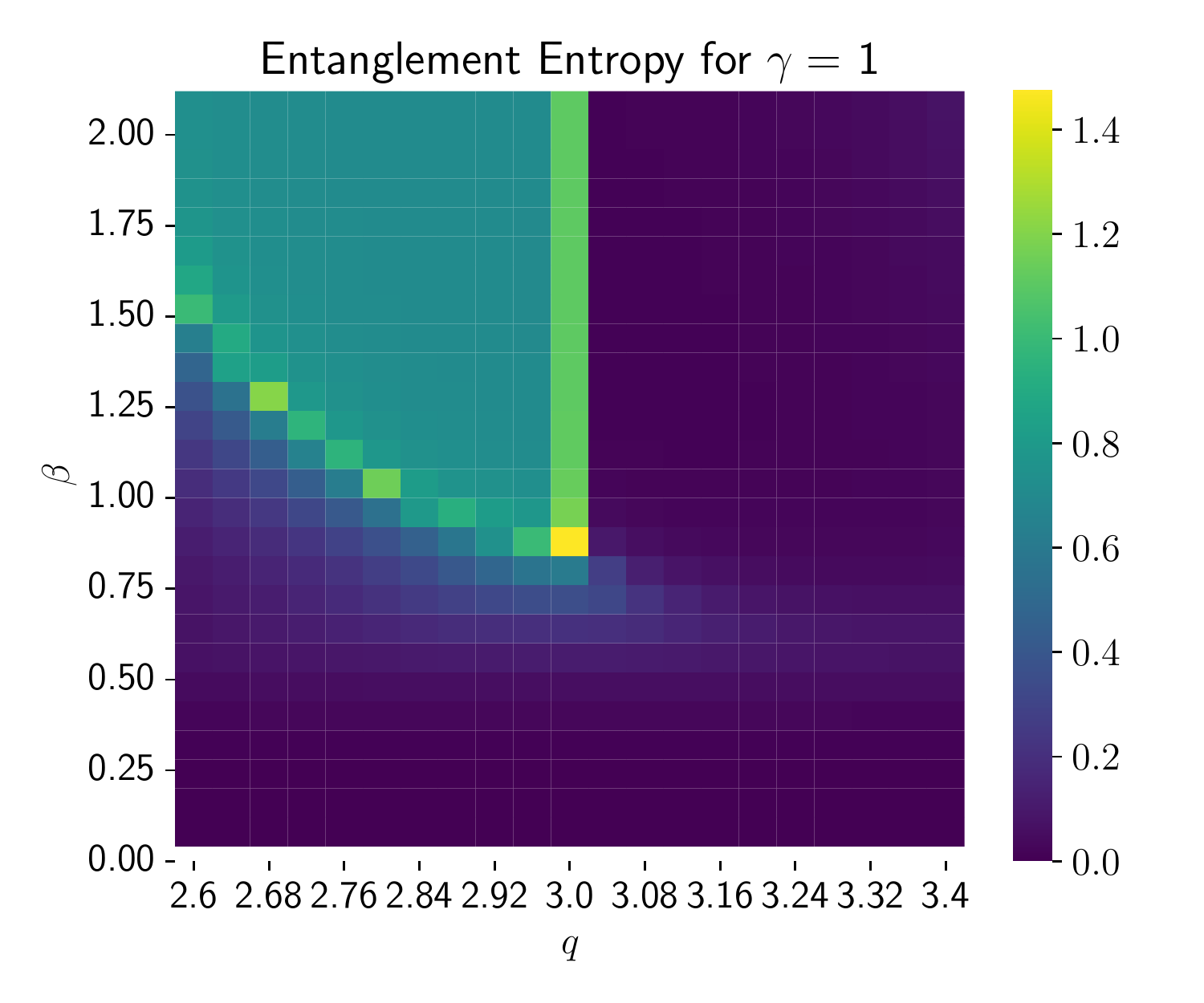}
\end{subfigure}
\caption{Here, we look more closely at $\gamma=1$ in the neighborhood of $q=3$. In the left panel, we have the specific heat, which shows fairly smooth behavior even as one crosses an integer value of $q$. In the right panel, we look at the entanglement entropy, which shows that a discontinuity at integer values of $q$ persists to small values of $\gamma$. The entanglement entropy takes a value of $\ln 3$ along the line $q=3$ for sufficiently large $\beta$---implying a line of $\mathbb{Z}_3$ order along this line. Within the lobe region, the entanglement entropy takes a value of $\ln 2$---implying a region of $\mathbb{Z}_2$ order. \label{fig:nearq3}}
\end{figure}

In Figure~\ref{fig:nearq3}, we look more closely at $\gamma=1$ in the neighborhood of $q=3$. In the left panel, we have the specific heat, which shows fairly smooth behavior even as one crosses an integer value of $q$. In the right panel, we look at the entanglement entropy. From the heatmap of the entanglement entropy, we see that the discontinuity at $q=3$, which disappears for the specific heat as one dials $\gamma$ toward zero, persists however for the entanglement entropy. In fact, the entanglement entropy takes a value of $\ln 3$ along the line $q=3$ for sufficiently large $\beta$---implying a line of $\mathbb{Z}_3$ order along this line. Within the lobe region, the entanglement entropy takes a value of $\ln 2$---implying a region of $\mathbb{Z}_2$ order. Both of these are consistent with the $\gamma=\infty$ phase diagram, which suggests that the $\gamma=\infty$ phase diagram might persist to all finite values of $\gamma$, despite the smooth lines seen in the heatmaps of the specific heat.

\section{Summary and Ongoing Work}

We are studying the O(2) model extended with a symmetry-breaking term $\gamma \cos(q\varphi)$. The model has two continuously tunable parameters, which allows us to tune from the Ising model, through the $q$-state clock models, and to the $XY$ model by varying $q$ from $q=2$ to $q\rightarrow \infty$, and it allows us to tune from the XY model to a given $q$-state clock model by varying $\gamma$ from $\gamma=0$ to $\gamma=\infty$.

Previously, we established the $\gamma=\infty$ plane of the 3 parameter ($q$, $\beta$, $\gamma$) phase diagram. When $q$ is an integer, we recover the classic $q$-state clock model which has a single second-order phase transition for $q=2,3,4$ and two BKT transitions for $q \geq 5$. When $q$ is noninteger, we get a crossover and a second-order phase transition. Between integer values of $q$, the phase transition lines are smooth and continuous, however, across integer values of $q$ they are not. For example, between $3<q\leq 4$, there is a continuous second-order transition line and similarly for $4<q\leq 5$. However, there is a discontinuity at $q=4$. On the other hand, at $\gamma=0$, there is a single continuous line of BKT transitions, corresponding to the $XY$ model for all values of $q$. The question we are faced with is how does the $\gamma=\infty$ phase diagram connect to the $\gamma=0$ phase diagram as $\gamma$ is varied? Does the $\gamma=\infty$ phase diagram persist to all finite values of $\gamma$, or is there some value of $\gamma>0$ at which the phase transition lines are smooth and continuous?

In this work, we perform a preliminary exploration of the finite-$\gamma$ region of the phase diagram. The specific heat, which develops a peak near values of $\beta$ corresponding to crossovers or phase transitions, can serve as a proxy for the phase diagram. As we tune to smaller values of $\gamma$, the discontinuities that exist at infinite $\gamma$ disappear and the lines become smooth and continuous---giving the possibility that smooth and continuous phase transition lines occur across integer values of $q$. However, if we look at the entanglement entropy, we see that the original discontinuities seem to persist---suggesting that the $\gamma=\infty$ phase diagram may persist to all $\gamma>0$.

To rigorously characterize the phase diagram at finite $\gamma$, we are currently performing a finite size scaling study of this model to extract the critical exponents $\nu$, $\alpha$, $\beta$, $\gamma$, and $\eta$ using MCMC and TRG. We expect to have those results very soon and to give a detailed picture of the phase diagram at finite $\gamma$ in a forthcoming paper.

\acknowledgments

This work was supported in part by the U.S. Department of Energy (DOE) under Awards No. DE-SC0010113 and No. DE-SC0019139. The Monte Carlo simulations were performed at the Institute for Cyber-Enabled Research (ICER) at Michigan State University.


\providecommand{\href}[2]{#2}\begingroup\raggedright\endgroup

\end{document}